\documentclass[12pt]{article} 

\usepackage{graphicx} 
\usepackage{subfig} 
\usepackage{pdflscape} 
\usepackage{booktabs} 
\usepackage[left=2.5cm,right=3cm,top=3cm,bottom=2.5cm]{geometry}
\usepackage{fancyhdr} 
\usepackage[utf8]{inputenc}
\usepackage{float}
\usepackage{datetime}
\usepackage{natbib}
\usepackage{url}

\usepackage{setspace}
\usepackage{amsmath}
\usepackage{hyperref}
\usepackage{verbatim}
\usepackage{algorithm}
\usepackage{algpseudocode}
\usepackage{color}
\usepackage{footnote}
\usepackage{threeparttable}
\usepackage{caption}

\ddmmyyyydate

\author{Joshua Snoke, Gillian M Raab, Beata Nowok,\\
 Chris Dibben, and Aleksandra Slavkovic}
\title{General and specific utility measures for synthetic data}

\begin{document}
\maketitle

\begin{abstract}
Data holders can produce synthetic versions of datasets when concerns about potential disclosure restrict the availability of the original records. This paper is concerned with methods to judge whether such synthetic data have a distribution that is comparable to that of the original data, 
what we will term general utility.
We consider how general utility compares with specific utility, the similarity of results of analyses from the synthetic data and the original data. We adapt a previous general measure of data utility, the propensity score mean-squared-error ($pMSE$), to the specific case of synthetic data and derive its distribution for the case when the correct synthesis model is used to create the synthetic data. Our asymptotic results are confirmed by a simulation study. We also consider two specific utility measures, confidence interval overlap and standardized difference in summary statistics, which we compare with the general utility results. We present two examples examining this comparison of general and specific utility to real data syntheses and make recommendations for their use
for evaluating synthetic data.
\end{abstract}
\section{Introduction}
Dissemination of data to external researchers is an important goal for statistical agencies. With sensitive data,  the agencies may be constrained in their ability to allow access to raw records, except perhaps to approved users in restricted locations, such as data safe havens (e.g., U.S. Census Research Data Centers). To make their data more available agencies have developed methods of statistical disclosure control (SDC), also known as statistical disclosure limitation (SDL). SDC methods alter the data in order to reduce the risk of disclosure for sensitive information, i.e., protect privacy, while maintaining the utility of the data as judged by the validity of inference carried out using the altered data. Traditional methods include microaggregation, top or bottom coding, perturbation by adding random noise and the  swapping of values (e.g., for more details see \cite{fienberg2011data} and \cite{hundepool2012statistical}). 
 
An alternative SDC method involves the generation of synthetic datasets where some or all of the observed data have been replaced by synthetic values generated from models based on the original data. The risk of disclosure is reduced by replacing the original sensitive values. There is an extensive literature on methods for generating synthetic data and making inferences from them, e.g.,  \cite{RRR_2003}, \cite{reiter_partial}, \cite{kinn_R2010}, \cite{slavkovic10}, \cite{Drechsler_book}, \cite{synthpop_JPC}, to cite only a few key references. The U.S. Census Bureau, in partnership with academics, has made significant advances in practical applications and released several synthetic data products, including the Survey of Income and Program Participation (SIPP) Synthetic Beta Data \citep{SIPP}, the Synthetic Longitudinal Business Database \citep{KinneyEtAl2011} and \cite{OnTheMap}, a web-based interface to a partially synthetic version of the Longitudinal Employer-Household Dynamics dataset. Synthetic data are now becoming more widely accepted and are being developed by other institutions worldwide. For example, bespoke synthetic data are provided to individual users of the Scottish Longitudinal Study (SLSs), and the \textit{synthpop} package for R (\cite{Rsoft}) has been developed by \cite{synthpop_jss} to facilitate the generation of synthetic data extracts. Synthetic microdata are, however, still experimental and for the examples mentioned above they are supplied to the users to carry out exploratory analyses, but the final results for publication are almost always obtained from the original data. This final analysis is referred to as the gold-standard analysis.

It is well understood that inferences from synthetic data will only be valid if the models used to synthesize the data correspond to those that can be considered as having generated the original data. It is important for staff synthesizing the data to assess how well this condition is fulfilled by their synthetic dataset, and this can be done by so-called \emph{general} and \emph{specific} measures of utility. The former are summaries of differences between the distributions of the original and the altered data while the latter compare the differences between results from particular analyses. 

Synthetic data utility has most often been assessed by \emph{analysis-specific measures} which compare data summaries and/or the coefficients of models fitted to synthetic data with those from the original data. If inferences from original and synthetic data agree, the synthetic data are said to have high utility. Published evaluations of synthetic data using specific utility measures, usually for just a few selected analyses, have highlighted differences in the quality of syntheses (\cite{Reiter_RSS}, \cite{RiskIAB}, \cite{KinneyEtAl2011}, \cite{miranda2016using}, \cite{Nowok2015}). However, when an agency prepares synthetic data for a user they will not know, except in very general terms, what analyses will be carried out. In practice, a user usually carries out a number of exploratory analyses in order to decide which models to fit and present. 
When the synthesizer does have some knowledge of the models that the analyst has in mind and bases the synthesis on these models, this may falsely reassure the analyst that their model is the correct one. When the generative model that informs the synthesis adheres too closely to the proposed utility model, the validity checks such as the existence of other interactions will not be apparent in the synthesized data; see \cite{NowokRaabDibbenSJIAOS} and  \cite{synthpop_JPC} for examples. Thus \emph{general measures} of utility could be more helpful in allowing an assessment of how well the final inference might agree with what would have been obtained had the user had access to the unchanged data for all of the analyses, rather than just at the final stage of a gold-standard analysis. Global measures of utility that can be used for any type of altered data have been proposed by several authors, such as \cite{KarrUtil06} and \cite{WooProp}, and \cite{Drechsler_book} has illustrated their use for a real example of synthetic data.

The disclosure risk associated with releasing any data from a statistical agency is clearly important. Agencies most commonly release data in the form of cross-tabulations or other summaries and there is an extensive literature on methods for assessing disclosure risk for such data, see \cite{WandDeW2001} for a review. Disclosure risk measures for micro-data releases, such as synthetic data, are less well developed. Methods have often been tailored to individual data products, e.g. \cite{ElliotSYLLS}, \cite{drbmemnov2006}, \cite{RiskIAB}, \cite{Loongetal_2013}. More recent research with synthesized categorical data has proposed methods that can be used to identify individual records with disclosure potential (e.g., \cite{Hu_etal,ReiWangZhang,Mcclure_reiter_2015}), but at present does not provide measures that can be used with the complexity of real data sets. This is clearly an area where further research is required but we do not address it here where our focus is on utility measures.

In this paper we evaluate and recommend extensions to existing global and specific measures of utility for the special case of synthetic data, and we compare general utility results to specific utility for data generated by different methods of synthesis. In Section 2 we review methods for generating and making inference from synthetic data and introduce our notation. In Section 3 we review previous work on general utility measures. In Section 4 we extend previous work for a propensity-score-based general utility measure by proposing two statistics specifically designed for synthetic data. In Section 5 we cover specific utility measures typically used for synthetic data. In Section 6 we give two data examples comparing outcomes based on general and specific measures and highlighting differences in their evaluation of different syntheses. In Section 7 we offer concluding remarks and recommendations.

\section{Brief Review of Synthetic Data Methodology}
\subsection{Data synthesis}
Synthesis is performed by a researcher with access to the original data which we denote as ($X, Y$), where $X$ denotes the data that will be released with their original values and $Y$ are the sensitive data that will be replaced with synthetic values. It is possible for all the data to be synthesized in which case $X$ is empty. The synthesizing process assumes that the data come from an underlying joint generative distribution, $f(Y | X, \theta)$.

Here we consider the situation when new values of $Y$ are generated by fitting the observed data to $f(Y | X, \theta)$ to give an estimate $\hat{\theta}$ and by generating a new sample from $f(Y | X, \hat{\theta}).$ In practice this is typically approximated with a sequence of conditional models. A total of $m$ synthetic datasets are produced, where $m=1$ gives just a single dataset.  Some methods of inference from synthetic data require that the synthetic data are generated from the posterior distribution of $Y$, given the observed data. In the next section we discuss why these requirements do not apply in our case.

\subsection{Inference for Synthetic Data}
Inference from synthetic data can be required for a particular model or a set of summary statistics defined by a parameter vector $Q$. Inference involves carrying out the same procedure on each of the synthetic datasets and using their average $\bar q_m = \frac{1}{m}\Sigma_{i=1}^m q_i$ as an estimate of $Q$, where $q_i$   denotes the estimate from the $i^{th}$  synthesis  and  $v_i$ the estimate of its variance. We are assuming throughout this paper that methods appropriate for simple random sampling are used for inference from both the original and synthetic data.

Much of the literature on synthetic data is concerned with using the synthetic data to make inferences to the population parameter $Q$, allowing for both the variation between $Q$ and its value in the samples of observed data and the differences between the original and synthetic data. However, for this paper we focus on the situation where researchers use synthetic data produced for exploratory purpose and then will carry out a gold-standard analysis (i.e., using the original data) after models are chosen. This scenario has been implemented by some researchers and agency staff; see \cite{NowokRaabDibbenSJIAOS} or \cite{VerificServ09} for further examples. When such a gold-standard analysis is to be carried out, the user of synthetic data is interested in approximating (via synthetic data) the estimate $\hat{Q}$ and its variance-covariance matrix  $V_{\hat{Q}}$ that would be obtained from the observed data.

When the original data are generated from the same model used for synthesis, and when the asymptotic conditions specified in \cite{RRR_2003} and \cite{synthpop_JPC}  are met, $\bar q_m$ is a consistent estimator of $\hat{Q}$, and the simple plug-in estimator $\bar v_m = \frac{1}{m}\Sigma_{i=1}^m v_i$ is a consistent estimator of $V_{\hat{Q}}$. Note that neither multiple syntheses with combining rules nor sampling from the posterior distribution of $Y$ are required to calculate these quantities, see \cite{synthpop_JPC}.

To evaluate specific utility,  we compare results from the synthetic datasets with what would be obtained from the original data. Thus we need not be concerned with population inference, and can compare confidence intervals and standardized coefficients from the original data with the equivalent quantities for synthetic data, calculated from $\bar q_m$ and $\bar v_m$.  This approach uses the same estimator for any type of synthetic data, e.g.,  whether all of the observations or only selected variables or data values are  synthesized. If the data generating model used for the synthetic data is the one that generated the original data then the confidence intervals from the synthetic data will be consistently estimated by this approach, see \cite{synthpop_JPC} for more on inference with synthetic data under different situations. 

\section{General Utility Measures for Masked Data}
Previous work has suggested various general measures of utility for data that have undergone disclosure control. Generally these measures consider the distributional similarity between the original and the masked datasets, with greater utility attributed to masked data that are more similar to the original data. In the broadest sense, measures such as distance between empirical CDFs or the Kullback-Liebler (KL) divergence give an estimate of difference. 

\cite{karr2006new} and the follow-up paper \cite{WooProp} discussed and implemented various distributional measures such as the KL divergence, an empirical CDF measure, a method based on clustering, and one that uses propensity scores to estimate general utility. They compared these measures for microaggregation, additive noise, swapping, and resampling methods, and they evaluated the propensity score method as the most promising.

In this paper we focus on expanding this measure for the specific case of synthetic data. Propensity scores represent probabilities of group memberships, commonly used in causal inference studies. To use them as a measure of utility, we need to model group membership between the original and the masked data to get an estimate of distinguishability. Small distinguishability relates to high distributional similarity between the original and masked data. If we can model the propensity scores well, this general measure should capture relationships among the data that methods such as the empirical CDF may miss.

The propensity score method, given in \cite{WooProp}, described in Algorithm \ref{alg1}, proceeds as follows. The $n_1$ rows of the original and $n_2$ rows of the masked datasets are combined with the addition of an indicator variable $I$ giving the source of the data (0 for original data and 1 for altered). A propensity score $\hat{p_i}$ is estimated for each of the $N=n_1+n_2$ rows, as the probability of classification for the indicator variable, using predictors based on the variables in the data. The mean squared difference between these estimated probabilities and the true proportion of records from the masked data in the combined data (denoted $c=n_2/N$, usually $\frac{1}{2}$), gives the {\em utility statistic} $\frac{1}{N}\Sigma({\hat{p}_i - c})^2$ (the propensity score mean-squared error, henceforth referred to as $pMSE$). In the case of synthetic data with $m>1$ the $pMSE$ would be
calculated for each dataset and the mean taken as the overall utility. The method can be thought of as a classification problem where the desired result is poor classification (50\% error rate), giving better utility for low values of the $pMSE$.
\begin{algorithm}
\caption{General Utility Statistic Based on Propensity Score Mean-Squared Error}
\label{alg1}
	\begin{algorithmic}[1]
		\item stack the original $n_1$ rows, $Y_{real}$ and the $n_2$ rows of masked data $Y_{syn}$ to create the  $N=n_1+n_2$ rows of  $Y_{comb}$
		\item add an indicator variable, $I$, to $Y_{comb}$ s.t. $I = \{1: y_i \in Y_{syn}\}$
		\item fit a model to predict $I$ using predictors $Z_{comb}$ calculated from $Y_{comb}$.
		\item predict propensity scores, $\hat p_i$, for each row of $Z_{comb}$ 
		\item obtain the utility statistic from $\frac{1}{N}\Sigma^N_{i=1}({\hat{p}_i - c})^2$ where $c=n_2/N$ is the proportion of records in $Y_{comb}$ from $Y_{syn}$
	\end{algorithmic}
\end{algorithm}

\section{General Utility for Synthetic Data}
We extend the propensity score method for general utility specifically for the case of synthetic data. In particular, when the $pMSE$ is calculated from a logistic regression, we derive its large-sample expectation and variance under the null case of synthesizing data from the correct generative model of the original data, and use this to standardize the observed $pMSE$. 

This standardization transforms to a scale that has a clear interpretation for synthetic data. The previous use of the propensity score measure for general utility gave better utility as the value became closer to 0, where a value of 0 would occur when the original and altered data are identical. This is highly unlikely to occur for synthetic data as the goal is not to have identical entries, but to achieve the distributional similarity between the distribution of the observed data and the model used to generate the synthetic data. This condition is required for any inferences from synthetic data to be valid, and we will refer to it as ``Correct Synthesis" or CS. With expressions for the expectation and standard deviation of the $pMSE$ for synthetic data under CS, we can use two standardized statistics either the ratio to its expectation under CS, the  $pMSE$ ratio, or the standardized $pMSE$ calculated as its difference from this expectation in units of the standard deviation under CS. The former will have an expected value of $1$ under CS and the latter an expectation of zero and a standard deviation of $1$. In both cases, increased values of these statistics will be expected if CS does not hold.

We also consider other models used to compute the $pMSE$s, such as non-parametric classification and regression trees (CARTs) which may improve the specification of utility for complex datasets over previous used models such as logistic models, general additive models or polynomial splines. In this case the theoretical results for the null $pMSE$ do not hold, but we show null values can be approximated using resampling techniques. CART models were found to be promising for measuring utility in complex datasets and are included in the real data examples.

These general utility measures, with a choice of model for the propensity score, are implemented in the \emph{synthpop} package \citep{synthpop_jss}, so data custodians creating synthetic data will be able to compute the $pMSE$, $pMSE$ ratio or the standardized $pMSE$ as measures of the appropriateness of different synthesis models.

\subsection{Null Distribution of the Mean-Squared Error calculated from a logistic regression}

We first consider the null distribution of the $pMSE$ under CS when all the data are synthesized and derive asymptotic expressions for the expectation and variance of this null $pMSE$. Using simulated data we show that these expressions are valid under CS and that $pMSE$ values grow further from their null expectation as the difference between the  models generating the original and synthetic data increases.

\subsubsection{Theoretical Results: The Null \textit{\textbf{pMSE}} Distribution}
\label{sec:theornull}
To obtain the $pMSE$ from a fully specified parametric model we use a set of predictor variables $Z_{orig}$ calculated from the original data and an equivalent set $Z_{syn}$ from the synthetic data. We assume that these predictor matrices have the column dimension $k$, including in each case a vector of 1's for the intercept term. Note that, for the distribution of the $pMSE$, $Z_{orig}$ is a fixed quantity and $Z_{syn}$ is a matrix of random variables generated by the synthesis process.

When CS applies $Z_{orig}$ is generated from a sample from $f(y|\theta)$ and $Z_{syn}$ from  $f(y|\hat{\theta})$, where $\hat{\theta}$ is estimated from the original data, we show in Appendix~\ref*{sec:app_null} that the null $pMSE$ is distributed as a multiple of a chi-squared distribution with $(k-1)$ degrees of freedom and expectation and standard deviation given by 
\begin{equation}
\label{eq3}
E[pMSE] = (k-1)(\frac{n_1}{N})^2(\frac{n_2}{N})/N = (k-1)(1-c)^2c/N,
\end{equation}
\begin{equation}
 \label{eq4}
 StDev(pMSE) = \sqrt{2(k-1)}(\frac{n_1}{N})^2(\frac{n_2}{N})/N = \sqrt{2(k-1)}(1-c)^2c/N,
\end{equation}
where $n_1$ is the number of observations in the original data, $n_2$ number of observations in the synthetic data, $N = n_1 + n_2$ and $c = n_2/N$. In the most common case when $n_1$ and $n_2$ are equal, the expectation becomes $(k-1)/(8N)$ and the standard deviation $\sqrt{2(k-1)}/(8N)$. The primary assumptions underlying these results are that the estimated propensity scores are not close to 0 or 1 and that the expectations for the synthetic variables under repeated syntheses will be the means of the original variables. These are discussed further in Appendix~\ref{sec:app_null}.


Appendix 9.3 discusses the distribution of the $pMSE$ calculated from two synthetic datasets, generated from the same original data, by the same method used to compare the synthetic data with the original ones. We discuss in Section 4.3.1 why a comparison of pairs of syntheses can be useful as a method of estimating the null $ pMSE $ distribution when it cannot be derived theoretically.

\subsection{Incompletely synthesized data}
When some part of the data are left unchanged this may involve synthesizing only selected variables (incomplete by variables), only selected records (incomplete by rows), or only some variables for some observations (incomplete by observations). When synthesis is incomplete by rows or by observations, the selection is usually restricted to those observations that are expected to pose a high disclosure risk such as observations with extreme, potentially disclosive, values. When this is the case, estimation of the models used to create the synthetic data must use records from only those observations that will be replaced \cite{reiter_partial}. Our theoretical results will not apply because the selected observations will not follow the same distribution as the complete data. This will also be the case even for randomly selected rows, unless the $pMSE$ is calculated from only the synthesized rows. 

Our results are easily extended to the case of incomplete variables, see Appendix~\ref{sec:app_part}. In that case, the contribution from predictors depending only on unsynthesized columns is zero, since all values are unchanged. Equations~(\ref{eq3}) and (\ref{eq4}) still hold  with $k$ replaced by $k^*$ the number of variables in the predictor matrix which relate to synthesized variables (including interaction terms between synthesized and unsynthesized variables). The following section presents simulation studies confirming these results, both for complete and incomplete synthesis, with a multivariate Normal example. The simulation also illustrates the behaviour of the $pMSE$ ratio, or the standardized $pMSE$ under increasingly incorrect synthesis.

\subsection{Simulation to Validate Asymptotic Expressions for the Expectation and Variance of \textit{\textbf{pMSE}}}
\label{sec:app_sim1}
We present simulation studies to show that the asymptotic results derived in Appendix~\ref{sec:app_null} and \ref{sec:app_part}  hold under CS and to show how they deviate from the expectations for incorrect synthesis. We ran 1000 simulations, and for each simulation we generated ten original datasets (referred to henceforth as $Real$ datasets) of size 5000 from a multivariate Normal distribution of dimension 10 with means 0, variances 1, and off-diagonal covariances of the $i_{th}$ dataset taking values 0, 0.1, ..., 0.9 for $i=1, ..., 10$. 

In the first simulation, for each $Real$ dataset we generated a correct and incorrect complete synthesis. For the correct synthesis we use the variance matrix fitted to the $Real$ data to generate synthetic multivariate Normal data. For the incorrect synthesis we use the sample means and a  variance matrix with its off-diagonal elements set to 0. The incorrect synthesis uses a model progressively further from the true generative model as the $Real$ data are generated from a model with covariances that increase from 0 to 0.9.  This emulates synthesis that fails to account for correlations between the variables.

We model the propensity scores with a logistic regression model including all main effects and first-order interactions for the variables, but omitting the quadratic terms, giving us $k= 56$ parameters. The expected mean of the $pMSE$ under CS becomes:

\begin{equation} 
\label{eq2}
E[pMSE]= (k-1)(1-c)^2c/N = 55*0.5^3/10000 = 0.000688
\end{equation}
 and its standard deviation:
 \begin{equation} 
 \label{eq5}
 StDev(pMSE)= \sqrt{(2(k-1)}(1-c)^2c/N = \sqrt{110}*0.5^3/10000 = 0.000131
 \end{equation}

Table \ref{tab:null-sim1} gives the means of the simulation results. For correct synthesis the mean $pMSE$ agrees with equation~(\ref{eq2}) and that of its standard deviation with equation ~(\ref{eq5}) (data not shown for ~(\ref{eq5})). Thus the $pMSE$ ratio (mean $pMSE$ divided by ~(\ref{eq2})) and the standardized $pMSE$ (mean $pMSE$ minus ~(\ref{eq2}) divided by ~(\ref{eq5})) are close to 1 and 0 respectively, as expected. Values below 1 for the ratio $pMSE$ or 0 for the standardized $pMSE$ are acceptable, simply a result of random variation, and implying correct synthesis.

For the incorrect syntheses models that fail to capture the correlations between the variables, $pMSE$ values compared to the original data increase as the covariance values increase as does its standard deviation (the latter not shown). Note that for the first line of Table \ref{tab:null-sim1} when the synthetic data are generated from a model with covariances of zero, it still does not give a value at the expectation, as was the case for synthesis from the correct model. This is because even though the population covariances are set to 0, the simulated $Real$ data do not have exactly zero covariances, so the incorrect synthesis here is not generated from a model correctly fitted to the observed data. As the covariances in the original data increase the $pMSE$ ratio and the standardized $pMSE$ increase, the latter very steeply. The ratio is an appropriate measure of the discrepancy which the $pMSE$ model finds between the two distributions. The standardized value gives a measure (like a t-statistic) of its deviation from the null value for CS. Given that we know that CS can rarely be fully achieved, except for simulated data, the standardized measure may be over-sensitive to small differences and  the ratio $pMSE$ is likely to be a more useful measure.
%
%
 \begin{table}[ht]
 	   \caption{\label{tab01} Results from 1,000 simulated syntheses of multivariate Normal data using correct and incorrect models with the $pMSE$ calculated from a logistic model including all main effects and first order interactions.}
 	\centering
 	\begin{tabular}{cccccccccccc}
 		\hline
 		Population & \vline & \multicolumn{3}{c}{ Correct Synthesis } & \multicolumn{3}{c}{Incorrect Synthesis } \\ 	
 		covariances & \vline & \multicolumn{3}{c}{$pMSE$} & \multicolumn{3}{c}{$pMSE$} \\
 		& \vline & Mean & Ratio & Standardized & Mean &  Ratio & Standardized\\ 
 		\hline
 		0.0 & \vline & 0.000684 & 0.995 & -0.024 & 0.00124 & 1.805 & 4.221 \\ 
  		0.1 & \vline & 0.000693 & 1.007 & 0.039 & 0.01428 & 20.77 & 103.7 \\ 
  		0.2 & \vline & 0.000696 & 1.013 & 0.068 & 0.03158 & 45.93 & 235.6 \\ 
  		0.3 & \vline & 0.000688 & 1.000 & 0.001 & 0.04696 & 68.31 & 353.0 \\ 
		0.4 & \vline & 0.000686 & 0.998 & -0.008 & 0.06021 & 87.57 & 454.0 \\ 
		0.5 & \vline & 0.000686 & 0.998 & -0.010 & 0.07202 & 104.8 & 544.1 \\ 
		0.6 & \vline & 0.000684 & 0.996 & -0.024 & 0.08248 & 120.0 & 623.9 \\ 
		0.7 & \vline & 0.000686 & 0.998 & -0.010 & 0.09192 & 133.7 & 695.9 \\ 
		0.8 & \vline & 0.000688 & 1.001 & 0.005 & 0.10054 & 146.2 & 761.7 \\ 
		0.9 & \vline & 0.000691 & 1.005 & 0.029 & 0.10830 & 157.5 & 820.9 \\ 
		\hline
		 	\end{tabular}\label{tab:null-sim1}
   	\begin{tablenotes}
 	\footnotesize
 	\item Ratios and standardized scores from theoretical expectations.
 \end{tablenotes}		

 \end{table}

In the second simulation, for each $Real$ dataset we generated a correct and incorrect incomplete synthesis, leaving eight of the ten original variables unchanged. For the correct synthesis we fit linear models using all unsynthesized variables as predictors (and the first synthesized variable as a predictor for the second) to generate new synthetic draws. For the incorrect synthesis we take a parametric bootstrap of the two variables using the sample means and standard deviations. In the same way as the complete synthesis, the incorrect synthesis ignores the correlations between variables and grows progressively further from the true generative model as the $Real$ data are generated from a model with covariances that increase from 0 to 0.9.

Equations~(\ref{eq8}) and (\ref{eq9}) gives the new expected value and standard deviation of the $ pMSE $ with only two synthesized variables. Recall, $k^*$ is the dimension of the propensity score predictor matrix that involves synthesized variables. Including main effects and first order interactions, this reduces from 55 previously to 19. The simulation results given in Table \ref{tabA1} confirm this, as well as showing a similar pattern for the ratio and standardized $ pMSE $ values for incorrect synthesis as was seen in Table \ref{tab01}.

\begin{equation} 
\label{eq8}
E[pMSE]= (k^*-1)(1-c)^2c/N = 19*0.5^3/10000 = 0.0002375
\end{equation}
 and its standard deviat{}ion:
 \begin{equation} 
 \label{eq9}
 StDev(pMSE)= \sqrt{(2(k^*-1)}(1-c)^2c/N = \sqrt{38}*0.5^3/10000 = .000077055
\end{equation}
%
%
\begin{table}[ht]
\caption{\label{tabA1}
	 Results from 1,000 simulated syntheses of multivariate Normal data with only two of the 10 columns synthesized,  using correct and incorrect models with the $pMSE$ calculated from a logistic model including all main effects and first order interactions.
}
\centering
\begin{tabular}{rrrrrrrr}
  \hline
 Population & \vline & \multicolumn{3}{c}{ Correct Synthesis } & \multicolumn{3}{c}{Incorrect Synthesis } \\ 	
 		Covariance & \vline & \multicolumn{3}{c}{$pMSE$} & \multicolumn{3}{c}{$pMSE$} \\
 		& \vline & Mean & Ratio & Standardized. & Mean &  Ratio & Standardized
 		\\ 
  \hline
  0.0 & \vline & 0.000244 & 1.027 & 0.083 & 0.00045 & 1.902 & 2.781 \\ 
  0.1 & \vline & 0.000239 & 1.007 & 0.022 & 0.00618 & 26.00 & 77.05 \\ 
  0.2 & \vline & 0.000239 & 1.007 & 0.022 & 0.01553 & 65.39 & 198.5 \\ 
  0.3 & \vline & 0.000237 & 0.996 & -0.013 & 0.02551 & 107.4 & 328.0 \\ 
  0.4 & \vline & 0.000232 & 0.975 & -0.076 & 0.03563 & 150.0 & 459.3 \\ 
  0.5 & \vline & 0.000236 & 0.994 & -0.019 & 0.04576 & 192.7 & 590.8 \\ 
  0.6 & \vline & 0.000233 & 0.982 & -0.055 & 0.05614 & 236.4 & 725.5 \\ 
  0.7 & \vline & 0.000236 & 0.995 & -0.015 & 0.06697 & 282.0 & 866.0 \\ 
  0.8 & \vline & 0.000233 & 0.981 & -0.060 & 0.07849 & 330.5 & 1016 \\ 
  0.9 & \vline & 0.000232 & 0.978 & -0.066 & 0.09118 & 383.9 & 1180 \\ 
  \hline
\end{tabular}
  	\begin{tablenotes}
  	\footnotesize
  	\item Ratios and standardized scores from theoretical expectations.
  	\end{tablenotes}

\end{table}

\subsubsection{Using Resampling Techniques for the Distribution of the \textit{\textbf{pMSE}}}
\label{sec:resamp}
We can use the results above when calculating propensity scores using a fully specified logistic model which provides a value of $k$ for the number of fitted parameters, but we may be interested in using non-parametric models such as adaptive models, stepwise regression or CART. In these cases, we cannot use the previous results, but we would still like to estimate the null $pMSE$. We show that the null distribution can be estimated using resampling techniques. The theoretical derivations in the previous section assumed that the two datasets compared were drawn from the same underlying generative model. By resampling, we can compare two datasets which we know were generated from identical distributions, and we can use the resulting $pMSE$ values as an estimate of the theoretical null $pMSE$.

One such resampling method is to calculate the $pMSE$ between pairs of synthetic datasets generated from the same original data. This estimates 
a $pMSE$ under CS even when the synthesizing model is incorrect, since both datasets are drawn from the same distribution. If a large enough number of pairs of syntheses are produced they can be used to estimate the mean and variance of the $pMSE$.
%
This method requires much additional computation if only one synthetic set is planned. An alternative method in the case of a single synthetic dataset is to use a permutation test to obtain null expectations. We describe it here for the case when the synthetic data has the same number of records as the original. The indicator variable used with the $Z$ matrix from the original and a single synthetic dataset is permuted, and a $pMSE$ calculated from each permutation (see Algorithm~\ref{alg2}). This method can be less computationally burdensome than producing extra syntheses, and it can also produce utility estimates when only a single synthetic dataset has been produced. Its disadvantage is that it does not give the correct null $pMSE$ unless all the data are synthesized. This can be understood by considering the contribution to the $pMSE$ from columns of $Z$ corresponding the unsynthesized data $X$. In calculating the $pMSE$ from the original data there will be no contribution from these columns because the difference in means will be zero (see Appendix~\ref{sec:app_null}). But the contribution will not be nothing with the permutation distribution because the permutation no longer treats $X$ as fixed. An alternative approach would be to omit any $X$ variables from the calculation of the $pMSE$, but this would be unsatisfactory  since it would not evaluate whether the relationships between $Y_{syn}$ and $X$ were maintained.

\begin{algorithm}
	\caption{Permutation Test for Null Mean and Standard Deviation Estimates}
	\label{alg2}
	\begin{algorithmic}[1]
		\If{$m$ $>$ 1 synthetic datasets}
		\State randomly assign a synthetic dataset for each permutation
		\EndIf
		\For{each permutation}
			\State randomly shuffle the group indicator variable $I$ to produce $I_p$
			\State follow algorithm \ref{alg1} using $I_p$ in place of $I$
			\State obtain $pMSE_{Perm_i}$ from the predicted propensity scores
		\EndFor
		\item return $\overline{pMSE_{Perm_i}}$ and $sd(pMSE_{Perm_i})$ for null mean and s.d. values
	\end{algorithmic}
\end{algorithm}
Thus we propose two resampling methods that can be used when methods, such as CART, without a known number of parameters are used to calculate the distribution of the $pMSE$ and derive the $pMSE$ ratio and the standardized $pMSE$ utility statistics. To confirm our results, the simulation study described above was extended to include our evaluation of the resampling method, and it is included in Appendix~\ref{app:resampSim}. 
For logistic models with known $k$, the resampling methods gave estimates of the null distribution of the $pMSE$ under CS that agreed with the theoretical results (data not shown). For CART propensity score models, where we do not know $k$, the expected values under permutation stayed constant across different syntheses as expected and the ratio of the $pMSE$ to the null expectation increased as the model used for synthesis was further from the correct one. We present results for complete synthesis and for incomplete columns. We also investigated the possibility of using resampling methods for the null distribution of the $pMSE$ for synthesis with selected rows. While the pairwise method gave satisfactory results for randomly selected incomplete rows, we have not investigated the important, but more complicated, situation when  the data to be replaced are selected according to their perceived disclosure potential.  

\begin{table}[ht]
\begin{threeparttable}
 	   \caption{\label{tab:nullRecc} Estimation Methods of the Null $pMSE$ for Different Synthesis and Propensity Score Model Scenarios}
 	\centering
 	\begin{tabular}{ccccc}
		\hline
		& \vline & \multicolumn{3}{c}{Propensity Score Model Type} \\
		& \vline & Logistic regression & \vline & CART \\
		\hline
		Complete Synthesis & \vline & Theoretical & \vline & Pairwise or permutation \\
		Incomplete (Columns) & \vline & Theoretical & \vline & Pairwise \\
		\hline
 	\end{tabular}
  \end{threeparttable}
\end{table}

Table \ref{tab:nullRecc} summarizes the applicable methods under different synthesis and propensity score model scenarios. If the $pMSE$ is calculated from a method with a known number of parameters, $k$, then the ratio and standardized measures can be calculated from equations~(\ref{eq3}) and (\ref{eq4}) for both complete and incomplete (by variables) synthesis. For complete synthesis with a model where $k$ is unknown, the pairwise or permutation methods can be used to obtain the ratio and standardized $pMSE$, and when only some of the variables are synthesized then the only method possible for CART models is the paired comparisons of multiple syntheses. 
%

\subsection{Choice of model for the propensity score}

As \cite{WooProp} have discussed, the choice of model for the propensity score is crucial to the way in which the $pMSE$ measures compare masking methods. \cite{WooProp} evaluated some different logistic regression propensity score models, and they found it was important to include higher order terms, including cubic terms, for the $pMSE$ to discriminate between methods such as incorrect simulation, adding random noise, and aggregation. However, their simulated data largely relied on inappropriate marginal distributions for the incorrect model. This type of inadequacy should be readily checked for synthetic data by visual comparisons of the real and synthetic data, as is done in the $synthpop$ package (\cite{Nowok2015}). For their real data example, \cite{WooProp} used a model with all main effects and first-order interactions between variables, where generalized additive models were used for the continuous variables. This approach would seem to be a useful starting point, although it might be more helpful to use the transformations that would normally be used in modelling continuous variables, rather than the additive models.

We consider expanding the propensity score models to include classification and regression tree models (CART) (\cite{Breiman1984}). These models have proved useful for generating synthetic data (\cite{reiter_cart}), and have been shown to out-perform other machine learning techniques (\cite{drechsler2011empirical}, \cite{Nowok2015}) for this purpose. Additionally, boosted tree models have been found to be useful for estimating propensity scores in causal inference applications, see \cite{mccaffrey2013tutorial}. The flexibility of these models suggests that for complex datasets, they may outperform logistic models in discerning between which syntheses performs best. In our real data examples we find that CART can improve estimation.

It is well known that CART models are subject to over-fitting and parameters can be set to control the complexity to prevent this.  This is not generally a problem for generating synthetic data but it can be when the $pMSE$ score is calculated, since a substantial proportion of the propensity scores may be close to zero or 1 even under data generated from a correct synthesis. This leaves little room for the $pMSE$ value to increase when an incorrect synthesis model is used, since the over-fit model picks up higher differences even when the synthetic data are drawn from the CS. It is important to check whether drastic overfitting is occurring, by looking at the propensity scores, and if necessary to adjust the tuning parameters. 

At the other extreme the parameters should be set to allow adequate discrimination. If the classification tree fails to perform any splits all estimated propensity scores will equal $0.5$ and the $pMSE$ will be zero. While you may argue this indicates good synthesis, it more likely means the tuning parameters for the decision tree were not set appropriately. Perfect indiscernibility between original and synthetic data is highly unlikely, and such a result would almost never occur using parametric propensity score models. If this does occur in practice, we recommend adjusting the tuning parameters to ensure the CART model can perform some discrimination. 

For the case of a simple synthetic dataset, logistic models with first-order interactions should be tried first. As the data become more complex, we recommend also fitting parametric models with higher order interactions (if computationally feasible) and CART models for comparison. The utility function in the $synthpop$ package currently includes both CART models as well as logistic models with interactions between variables up to an arbitrary order.
\section{Specific Utility Measures for Synthetic Data}
\label{sec:specific}
In contrast to the general utility approach, we can measure the utility of a synthetic dataset by assessing the similarity of results for specific analyses using both the original and synthetic data. For high utility we expect close similarity between the results for the same analysis calculated from the two different data sources. Most of the previous literature has used specific utility measures rather than general measures, which has been more commonly used for other types of disclosure-controlled data, e.g., produced by top-coding or micro-aggregation, rather than for synthesized data. \cite{KarrUtil06} and \cite{VerificServ09} refer to this type of utility as fidelity measures, since it provides the masked data users with a measure of trustworthiness for the analysis compared to the analysis on the unreleased data. 

The most common and understandable examples of analysis-specific measures compare estimated summary statistics or general linear model coefficients obtained from the original and masked data. The percentage overlap of confidence intervals, for each of the coefficients or summary statistics of interest, are calculated from the observed and masked data, e.g., \cite{KarrUtil06}, \cite{VerificServ09}, \cite{RiskIAB},  \cite{slavkovic10}, and \cite{WooSlav2015}. An interval-overlap measure, given in equation (\ref{eq:IO}), can then be calculated for each statistic of interest and summarized by the average, with a higher $IO$ corresponding to greater utility. Note that this measure is negative when there is no overlap and will decrease as the intervals move further apart. 
\begin{equation}
\label{eq:IO}
IO = 0.5\left[\frac{min(u_o,u_s)-max(l_o,l_s)}{u_o-l_o}+\frac{min(u_o,u_s)-max(l_o,l_s)}{u_s-l_s}  \right]
\end{equation}

The $IO$ measure has been extended by \cite{KarrUtil06} to a measure of ellipsoid overlap ($EO$) which uses an estimate of the overlap between the joint posterior distribution of all the parameters for the original and synthetic data. The $EO$ is a more satisfactory measure because it allows for the correlation between the parameter estimates. However, it is much more onerous to compute, the easiest method involving simulation, and may be less easily understood by those analyzing the data. 

An alternative summary of the differences in summary statistics is the standardized difference between the original estimate and the synthetic estimate calculated as $|\hat{\beta}_{orig}-\hat{\beta}_{syn}|/s.e.(\hat{\beta}_{orig})$, where $\hat{\beta}_{orig}$ and $\hat{\beta}_{syn}$ are the coefficients of the same model estimated from the real and synthetic data and  $s.e.(\hat{\beta}_{orig})$ is the estimated standard error of the coefficients from the original data. This measure was used in \cite{WooSlav2015} to test data that had undergone post randomization method (PRAM), and it is similar to the standardized bias suggested by \cite{Loongetal_2013}, which differs only by using the estimated standard error from the synthesized data.

For our examples we present both the confidence interval overlap and the standardized difference as measures of specific utility. These two related measures are implemented in the \emph{synthpop} package under the \emph{compare.fit.synds()} function, and can be used to compare results from synthetic data to a gold-standard analysis once a researcher's code is run on the original data.  For a model with many coefficients the $IO$ and the standardized difference can be summarized by their mean or their median and range or, more usually, displayed graphically.

\section{Data Examples}
\label{sec:realdata}
As discussed in previous sections, specific utility measures the inferential usefulness of a dataset for a given model. Close inferential results between original and synthetic data allow publicly released data to be useful as an exploratory tool, for teaching, and possibly even for publication. The general measure of utility can provide value both by relating to the specific utility measures, but also by giving a different angle when the specific utility results are misleading. It is impossible to compare every possible specific analyses researchers may wish to perform using the synthetic data, and in some cases, the specific utility may be misleading, particularly in the utility models are similar to models included in the synthesis. The general utility and specific utility should be used in tandem along with data visualizations and marginal distribution checks to aid synthetic data producers in determining which synthesis is best for release. We use two real data examples to illustrate the need for this holistic approach by comparing the general utility
measures to several specific models.

\subsection{Scottish Health Survey}
We use data from the 2013 Scottish Health Survey (SHeS), focusing specifically on the data used for the 2015 report on Mental Health and Wellbeing, see \cite{wilson2015scottish}. This report uses a subset of the SHeS dataset containing 8,721 observations on 15 variables covering demographic information, behavioral factors, and mental health indicators. Table \ref{tab06} gives a detailed summary of the data. 

\begin{table}[ht]
 	   \caption{\label{tab06} Summary of Data for Report on Mental Health and Wellbeing}
 	\centering
 	\begin{tabular}{ccc}
		\hline
		Variable & Label & Range\\
		\hline
		\hline
		Sex & $Sex $& Male = 1, Female = 2\\
		Age Group & $ag16g10$ & 7 categories, minimum = 16 \\
		Martial Status & $maritalg$ & 6 categories \\
		Parental Employment Type & $pnssec5$ & 7 categories \\
		Income Quintile & $eqv5$ & 6 categories \\
		In $15\%$ Most Deprived Area & $SIMD15\_12$ & 1 = No, 2 = Yes \\
		Economic Activity & $econac12$ & 6 categories \\
		Provides Caregiving & $RG17a$ & 5 categories \\
		Physical Activity Level & $adt10gpTW$ & 4 categories \\
		Servings of Fruits and Vegetables & $porftvg3$ & 3 categories \\
		Has Alcohol Dependence & $AUDIT20$ & 1 = No, 2 = No Answer, 3 = Possible \\
		Smoker Status & $cigst3$ & 1 = Current, 2 = Never, 3 = Ex \\
		COP Diagnosis & $COPDDoct$ & 1 = Yes, 2 = No \\
		WEMWBS Mental Health Score & $wemwbs$ & 1=``issues"  0=``standard" \\
		GHQ12 Mental Health Score & $ghq12scr$ & 1=``issues"  0=``standard" \\
		\hline
 	\end{tabular}
\end{table}

The study focused on mental health outcomes for males and females as measured by the two scores, the Warwick-Edinburgh Mental Wellbeing Scale (WEMWBS) and the General Health Questionnaire (GHQ12), while controlling for demographic and behavioral factors. The WEMWBS is derived from 14 questions concerning personal thoughts and feelings with self-reported answers. The GHQ12 entails 12 experiential questions, six positively worded and six negatively worded, with self-reported responses of the participants' level of agreement. Specifically the models estimated, which we replicate, were four logistic regression models, two for men and two for women with the two mental health indicators as the response variables. While these responses were originally continuous they were recoded to be binary variables, with 1 indicating a significantly above average level of mental health issues and 0 indicating standard values. Table \ref{tab07} summarizes the four models.

\begin{table}[ht]
 	   \caption{\label{tab07} Wellbeing Fitted Models}
 	\centering
 	\begin{tabular}{cccc}
		\hline
		Model & Sex & Response & Covariates\\
		\hline
		\hline
		(1) & Male & $wemwbs$ & $ag16g10$, $maritalg$, $SIMD15\_12$, $econac12$, $eqv5$, $RG17a$,\\ 
		& & &  $adt10gpTW$, $AUDIT20$, $cigst3$, $porftvg3$, $COPDDoct$ \\
		\hline
		(2) & Female & $wemwbs$ & $ag16g10$, $maritalg$, $SIMD15\_12$, $econac12$, $eqv5$, $RG17a$,\\ 
		& & &  $adt10gpTW$, $AUDIT20$, $cigst3$, $porftvg3$, $COPDDoct$ \\
		\hline
		(3) & Male & $ghq12scr$ & $ag16g10$, $maritalg$, $pnssec5$, $econac12$, $eqv5$, $RG17a$,\\ 
		& & &  $adt10gpTW$, $AUDIT20$, $cigst3$, $COPDDoct$ \\
		\hline
		(4) & Female & $ghq12scr$ & $ag16g10$, $maritalg$, $pnssec5$, $econac12$, $eqv5$, $RG17a$,\\ 
		& & &  $adt10gpTW$, $AUDIT20$, $cigst3$, $COPDDoct$ \\
		\hline
 	\end{tabular}
\end{table}

Synthesizing all observations and all variables, we create one synthetic dataset for each of three different methods: sequential parametric regression models, sequential non-parametric CART models, and simple random sampling (non-parametric bootstrap samples of each variable). The last of these fails to model any dependencies between the variables and so would be expected to perform poorly.  General utility is measured using $pMSE$ estimated from two logistic models one with only main effects and the second with all main effects and first order interactions and also with a CART model. The logistic models had 44 and 964 degrees of freedom, respectively. Ratio and standardized $pMSE$s are also estimated, using expressions  for the null expected mean and standard deviation for the logistic models and from permutations for the CART model.

For specific utility we estimated models (1)-(4) with both the original and synthetic data and calculate the two statistics given in Section~\ref{sec:specific}, i.e., confidence interval overlap and standardized differences in coefficient values. For both of these, the median across all covariates in the models is reported.  All utility results are summarised in Table \ref{tab08}. 

\begin{table}[ht]
 	   \caption{\label{tab08} SHeS General and Specific Utility Results; Comparing Synthesis Models with Different $pMSE$ Models}
 	\centering
 	\begin{tabular}{lllccc}
		\hline
		& General Utility & $pMSE$&\multicolumn{3}{c}{Methods for Synthesis}\\
		& Measure & Model & $Parametric$ & $CART$ & $Sampling$\\
		\hline
		\hline
		& $pMSE$ & logistic main effects  & 0.000384 & 0.000370 & 0.000420 \\
		& $pMSE$ & logistic interactions  & 0.00726 & 0.0161 & 0.1280 \\
		& $pMSE$ & CART & 0.0417 & 0.0372 & 0.114 \\\\
		
		& $pMSE$ Ratio & logistic main effects  & 1.19 & 1.15 & 1.30 \\
		& $pMSE$ Ratio & logistic interactions  & 1.05 & 2.33 & 18.5 \\
		& $pMSE$ Ratio & CART & 1.01 & 0.925 & 2.47 \\\\
		
				& Standardized $pMSE$ & logistic main effects  & 0.908 & 0.701 & 1.43 \\
				& Standardized $pMSE$ & logistic interactions  & 1.14 & 29.1 & 384 \\
				& Standardized $pMSE$ & CART & 0.459 & -2.59 & 53.5 \\\\
		\hline
			& {Specific Utility} & Fitted model\\
			\hline
				\hline
		& {Median C.I. Overlap} & Model (1) & .704 & .667 & .588 \\
		&  {Median Std. $\hat\beta$ Diff.} & Model (1) & 1.38 & 1.45 & 1.89 \\
		\hline
		& {Median C.I. Overlap} & Model (2) & .868 & .663 & .282 \\
		& {Median Std. $\hat\beta$ Diff.} & Model (2) & 0.485 & 1.59 & 2.80 \\
		\hline
		& {Median C.I. Overlap} & Model (3) & .822 & .752 & .534 \\
		& {Median Std. $\hat\beta$ Diff.} & Model (3) & 0.792 & 1.39 & 1.99 \\
		\hline
		& {Median C.I. Overlap} & Model (4) & .815 & .612 & .482 \\
		& {Median Std. $\hat\beta$ Diff. }& Model (4) & 0.855 & 1.81 & 2.23 \\
		\hline
	\end{tabular}
\end{table}

\begin{figure}[h]
\centering
\includegraphics[width = 16cm, height = 12cm]{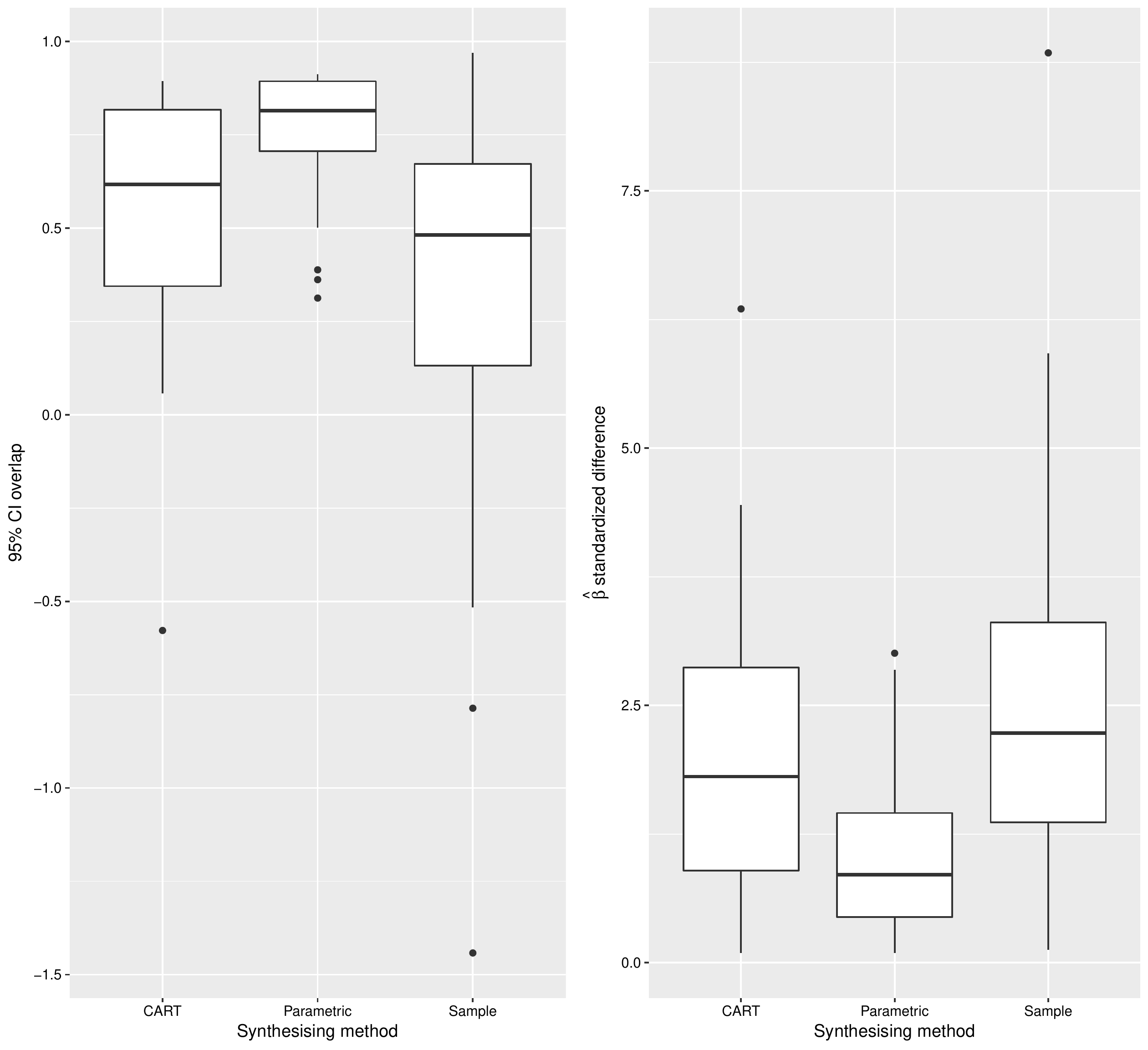}
\caption{\label{fig1} SHeS Model (4) Specific Utility}
\end{figure}
Considering the general utility results first, we can see that the logistic $pMSE$ model with only main effects does not discriminate between the synthesis methods and finds little evidence of a difference between the original or synthetic data (ratios close to 1). The logistic $pMSE$ model with interactions finds the sampling method quite inadequate. It judges the parametric model as being closer to the original data than the CART synthesis model. The rescaled $pMSE$ values calculated from CART models discriminate less between the first two synthesis methods than does the logistic with interactions but more so than the logistic with only main effects. For the CART $pMSE$ method, the CART and parametric synthesis methods give similar ratios, while the sampling is worse but not by as large a magnitude as the logistic $pMSE$ model with interactions. 

Looking at the specific utility results we can see that for all the fitted models the sampling utility is poorer than the other synthesis methods.  The specific utility for fitting model (1) is similar for parametric and CART syntheses, but the other three models show better specific utility for the parametric models. Model (4) has the largest differences between specific utility for different synthesising models and this is illustrated in Fig.~\ref{fig1} where  C.I. overlap and standardized differences for all 39 coefficients are displayed as boxplots. Thus, for three out of the four models the ranking of synthesis methods is similar to the general utility using either the logistic with interactions or CART propensity score models.

\subsection{Historical Census Data}

Our second example uses data from the 1901 Census of Scotland made available by the Integrated Census Microdata (I-CeM) project\footnote{https://www.essex.ac.uk/history/research/icem/}. These datasets have many features that make them similar to current census data from the UK, such as large sample sizes, mainly categorical data (some with small categories) and data organised by household, but they have the advantage that the original data are freely available to disseminate.

To illustrate our methods here we use a subset of the data consisting of private households in the historic county of Midlothian and the parish of the City of Edinburgh. The variables shown in Table~\ref{tabA2} consist of individual characteristics of the head of the household, plus data on family composition and number of rooms. The subset used was created by taking observations from the Edinburgh parish and countries of birth in the set $\{Scotland, England, Ireland, Germany, Italy, Russia, Wales\}$ with missing data on any variables removed, leaving 40,857 records. The variable $disability$ had a large number of categories, many with only a few cases. The largest category was ``Idiot or insane" reported by only 285 people.  Thus this category was reduced to a binary one indicating any disability. The variable $work\_status$ was derived from the census data on employer status and occupation and had three categories according as whether a head of household was a worker, an employer or this was irrelevant (e.g. students or retired people). This dataset is available in the [supplementary material] for this paper, as is the code used to create the synthetic datasets.
\begin{table} 
	\caption{\label{tabA2} I-CeM 1901 Historical Census Data}
	\centering
	\begin{tabular}{ccc}
		\hline
		Variable & Label & Range\\
		\hline
		\hline
		\emph{ Head of household characteristics } & & \\
		Sex & $sex$ & 2 categories \\
		Disability & $disability$ & 2 categories \\
		Martial Status & $mar\_stat$ & 5 categories \\
		Age (years) & $age$ & 9 - 96 \\
		Working Status & $work\_status$ & 4 categories \\
		\vspace{1mm} 
		Country of Birth & $ctry\_bth$ & 7 categories \\
		\emph{ Household characteristics } & & \\
		Number of Related Individuals & $n\_relations$ & 1 - 12 \\
		Number of Lodgers and Boarders & $n\_lodgers$ & 0 - 11 \\
		Number of Others (servants, visitors or unknown) & $n\_others$ & 0 - 26 \\
		Total Rooms in Dwelling & $totrooms$ & 1 - 54 \\
		\hline
	\end{tabular}
\end{table}

The variables $sex$ and $age$ are left unsynthesized. The remaining variables are synthesized using three different methods, CART, parametric models with the rank method (rank), and parametric models with normal linear regression rather than rank (normal). The rank method fits linear models to transformations of the numeric variables to z-scores, while the normal fits the models to untransformed data. The synthesis was conditional on the values of the unsynthesized variables and the order or the remaining variables, chosen by synthesizing the numerical variables first and the categorical variables second, was given by:
\[
\{totrooms, n\_others, n\_lodgers, n\_relations, work\_status, disability, mar\_stat, ctry\_bth\}
\]
15 synthetic datasets were generated from each method, and the observed $pMSE$ was taken as the average from the $pMSE$ calculated with each dataset. The null $pMSE$ was calculated using the average of the 105 pairwise comparisons of the 15 synthetic datasets in the case of the CART propensity models. Table \ref{tabI2} gives the observed $pMSE$ for each along with the two measures rescaling by the null of the two propensity score models.
\begin{table}[ht]
 	   \caption{\label{tabI2} I-CeM General and Specific Utility Results}
 	\centering
 	\begin{tabular}{lllccc}
		\hline
		& General Utility & $pMSE$ Model&\multicolumn{3}{c}{Methods for synthesis}\\
		& Measure  & &   &\multicolumn{2}{c}{$Parametric$}\\
		&  &  & $CART$ & $Rank$ & $Normal$\\
		\hline
		\hline
		& $pMSE$ & logistic interactions  & 0.001914 & 0.01567 & 0.002758 \\
		& $pMSE$ & CART & 0.009082 & 0.03078 & 0.1183 \\\\
		
		& $pMSE$ Ratio & logistic interactions  & 7.913 & 67.26 & 11.40 \\
		& $pMSE$ Ratio & CART & 3.557 & 11.24 & 33.90 \\\\
		
		& Standardized $pMSE$ & logistic interactions  & 64.98 & 611.3 & 97.76 \\
		& Standardized $pMSE$ & CART & 7.877 & 30.81 & 107.8 \\\\
		\hline
			& {Specific Utility} & Fitted model\\
			\hline
				\hline
		& {Median C.I. Overlap} & Model (1) & 0.3214 & 0.8429 & 0.8198 \\
		&  {Median Std. $\hat\beta$ Diff.} & Model (1) & 10.33 & 1.962 & 2.780 \\
		\hline
		& {Median C.I. Overlap} & Model (2) & 0.4519 & 0.9538 & 0.8718 \\
		& {Median Std. $\hat\beta$ Diff.} & Model (2) & 7.589 & 0.5604 & 1.959 \\
		\hline
		& {Median C.I. Overlap} & Model (3) & 0.6844 & -0.5306 & 0.5740 \\
		& {Median Std. $\hat\beta$ Diff.} & Model (3) & 4.776 & 18.05 & 6.462 \\
		\hline
	\end{tabular}
\end{table}

There is some disagreement between the assessments from the CART and logistic propensity score models, though they did agree in the sense that neither the CART or logistic propensity models rate any of the synthesis methods particularly well (no ratios close to 1). They both rank the CART synthesis model as the best, but switch the two parametric methods. 
Histograms comparing the marginal distributions of the original and synthesized data are available in the [supplementary material] to this paper. In particular, they show that the numeric variables are all highly skewed and this is not well captured by the parametric models. Thus it was reasonable to expect that the CART synthesis would have the highest general utility. Importantly, the logistic propensity score model fails to rate the parametric (normal) method as poorly as it should considering that method produces impossible, negative, values for all the numeric variables. The CART propensity model picks up on this. 

The normal parametric method which produces negative values is not a sensible approach. Regardless of how it performs on specific models, an analyst who found impossible values, such as negative household size, during exploratory analyses would likely lose all confidence in the entire synthetic data set, as they should. The utility measure from a CART model rates the models with negative values as having poor utility, but a complete utility evaluation should always include direct comparisons of the marginal distributions, either by comparing data summaries or by visualisation, as we illustrate in the histograms from the \emph{synthpop} package that are available in the [supplementary material].

\begin{figure}[h]
\centering
\includegraphics[width = \textwidth]{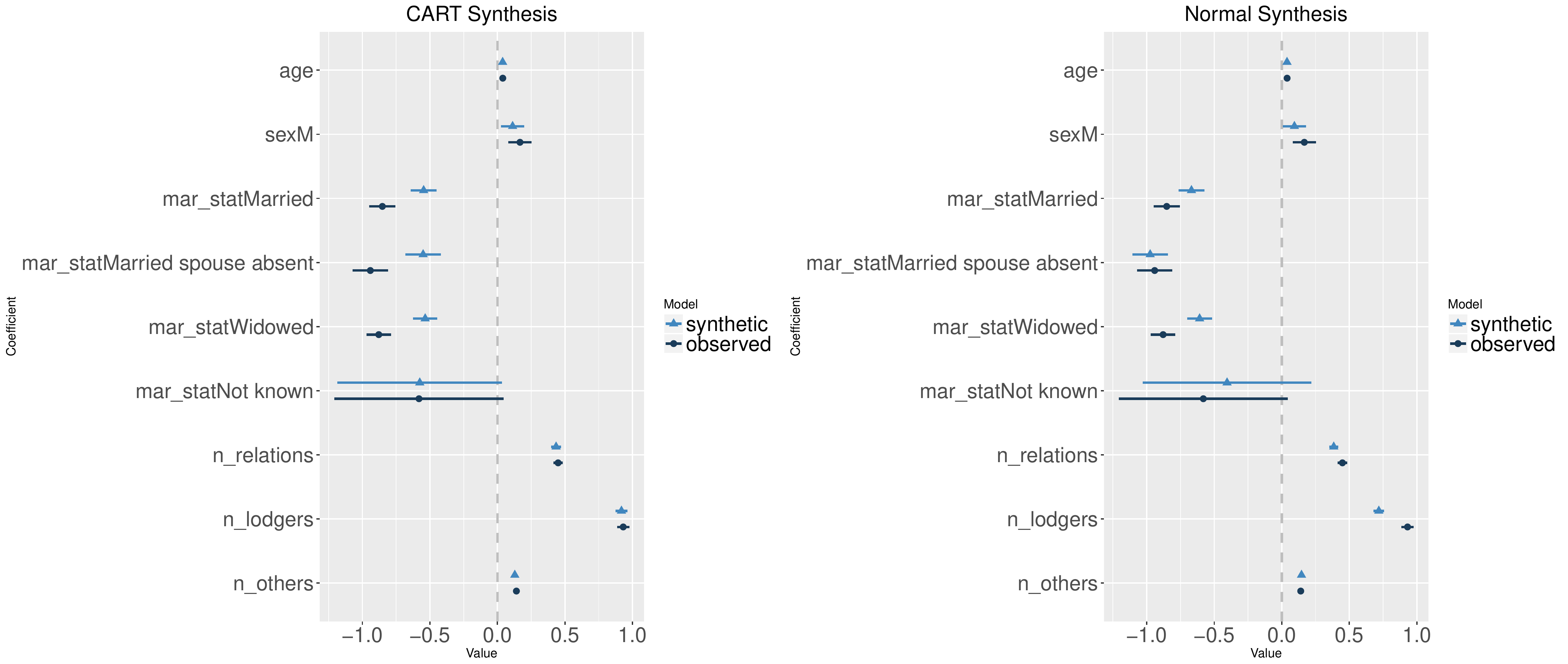}
\caption{\label{fig2} Model (2) C.I.s for CART and normal parametric synthesis}
\end{figure}

For specific utility, three models are estimated as shown in Table \ref{tabI1}, and measures of confidence interval overlap and standardized $\hat\beta$ difference are calculated. Results for each are given in Table \ref{tabI2}. The CART synthesis performs quite badly compared to the parametric syntheses on models (1) and (2). It may be surprising that the CART synthesis performs poorly, but recall that none of the general utility measures ascribed particularly high utility to any of the synthesis methods. Figure \ref{fig2} visualizes the confidence interval overlaps, and we see that the CART performs poorly primarily with the $mar\_stat$ variable, while the others are fine.

Both models (1) and (2) are very close to models that were included in the parametric syntheses, so it is unsurprising to see the parametric methods record high specific utility. These relationships were essentially baked into the parametric synthetic data, so the confidence intervals should be similar. On the other hand, we see in model (3), a model that is not closely represented in the parametric synthesis, that both parametric (rank) and parametric (normal) methods have much lower specific utility.
\begin{table}[ht]
 	   \caption{\label{tabI1} I-CeM Fitted Models}
 	\centering
 	\begin{tabular}{cccc}
		\hline
		Model & Response & Covariates\\
		\hline
		\hline
		(1) & $work\_status$ & $sex$, $age$, $mar\_stat$, $ctry\_bth$, $n\_relations$, $n\_lodgers$, $n\_others$\\
		\hline
		(2) & $disability$ & $sex$, $age$, $mar\_stat$\\
		\hline
		(3) & $totrooms$ & $sex$, $age$, $mar\_stat$, $n\_relations$, $n\_lodgers$, $n\_others$\\
		\hline
 	\end{tabular}
\end{table}

This example shows how specific utility can be very misleading. The specific utility is artificially increased when testing models that are closely related to models included in the synthesis process, even though the parametric (normal) synthesis of variables $totrooms,\\ n\_relations, n\_lodgers, n\_others$ is completely wrong. This example shows that general utility measures should not simply be used to avoid testing many specific models, but general measures can improve measurement in situations when specific utility is misleading. The inclusion of impossible values should immediately disqualify this synthetic data as a release dataset, and we would like our utility statistics to reflect that. The specific utility may fail to pick up on this, and the $pMSE$ based on logistic propensity score models is also less sensitive to this. While only one example, this shows the potential value of the non-parametric propensity score models.

\section{Conclusions}
In this paper we develop extensions to general  utility measures for synthetic data.  Our extensions include two new general utility statistics, the  $pMSE$ ratio and standardized $pMSE$, calculated  by standardizing the statistic by its null expected value and standard deviation. Rescaling by the null statistic aids in interpretability of utility specifically for synthetic data. Rather than basing our distance measure from identical data matrices, the standardized distance measures evaluate whether the synthesizing model is the correct one for the original data. Our measures are easier to compare because they do not depend on sample size.  The proposed functions are being implemented and will be available in the \emph{synthpop} package for R.

As \cite{WooProp} have discussed, the choice of model used to calculate the $pMSE$ is crucial to its performance as a utility measure. We proposed extending the models suggested by including non-parametric CART models to estimate propensity score values. Our examples of both real and simulated data suggest that CART models may be useful, particularly in the case of complex data. Parametric models with higher order interactions deserve further exploration, but are often not computationally feasible with many categorical variables. One solution might be to identify a subset of variables and investigate how well general utility suggests that relationships between them are maintained.

We also present some comparisons of how general and specific utility measures evaluate synthesis methods. In the first example, these show reasonable concordance between the ranking of different methods. In the second example, the specific utility gave misleading results, and the general utility served to give a different perspective. More experience with further synthesized data will be needed before we can generalize results, but it is clear the general measures aid in producing a complete utility evaluation. We believe that these new utility measures and their easy access via the \emph{synthpop} package will further support acceptance and use of synthetic data by researchers.
All utility measures in this paper are being implemented in the \emph{synthpop} package in R. The general utility measures can be used by staff creating synthetic data extracts to tailor the methods used to synthesise a data set so as to provide users with synthetic data that will be fit for their purposes. The researchers can use the specific utility measures when they carry out a gold-standard analysis at the end of their projects to be reassured that the synthetic data used for exploratory analyses was not misleading. Thus, our methodological developments will help agency staff to produce useful synthetic data and thus widen access to the use of confidential data by researchers.

\section{Acknowledgements}
This work was supported in part by NSF grants Big Data Social Sciences IGERT DGE-1144860 to Pennsylvania State University, and BCS-0941553 and SES-1534433 to the Department of Statistics, Pennsylvania State University. Beata Nowok and Gillian Raab are funded by the UK Economic and Social Research Council's Administrative Data Research Centre – Scotland, Grant ES/L007487/1.

\section{Appendix I}
\subsection{Null distribution of the \textit{\textbf{pMSE}} calculated from a logistic model}
\label{sec:app_null}
We assume the original data  has $n_1$ observations and the synthesized data $n_2$. To compute the $pMSE$ we fit a logistic regression model of the indicator variable $I$  on an $n_1+n_2$ by $k$ matrix of predictors  $Z$ where 
\[
Z =
\begin{bmatrix}
Z_{orig}  \\
Z_{syn}  
\end{bmatrix}
,
I =
\begin{bmatrix}
I_1 \\
I_2 
\end{bmatrix}
\]
and $I_1$ is a $n_1$ vector of zeros, $I_2$ a $n_2$ vector of ones, $Z_{orig}$ is derived from the original data and  $Z_{syn}$  from the synthesized data.  Note, that the usual formulae for the standard errors of logistic regression will not apply here, since $I$ is fixed, and not random. The distribution of any statistic is derived from that of the random variables $Z_{syn}$, conditional on the observed values of $Z_{orig}$ and $I$. Note that the $Z$ matrix here will include a column of $1$s for the intercept and will usually contain the original $Y_{orig}$ and $Y_{syn}$ values as well as interaction and  product terms or other functions calculated from them. 

A logistic regression model can be fitted by updating the current estimate of the coefficient vector $\beta^*$ by iterative reweighted least squares.  The weights are $w^*=p^*(1-p^*)$ where $p^*$ is the current estimate of the fitted proportion and the dependent variable is $Z'\beta^*+W^{-1}(I-p^*)$ where $W^*$ is an $N \times N$ diagonal matrix with elements $w^*$ (\cite{MandN_1989}).  Once the fitting has converged to give estimates $\hat{\beta}$ we can write the estimated coefficients of the logistic regression as 
\begin{equation} 
\label{eq1}
\hat{\beta}=(Z'WZ)^{-1}Z'W[logit(\hat{p})+W^{-1}(I-\hat{p})] 
\end{equation}
where $\hat{p}$ is the vector of predicted probabilities for each row of $Z$, i.e., the propensity score, and $W$ is an $N \times N$ diagonal matrix  with $i$ element $w_i=\hat{p_i}(1-\hat{p_i})$. Thus at convergence $W^{-1}(I-\hat{p})$ becomes zero leading to a set of $k$ equations:
\begin{equation} 
\begin{bmatrix}
Z'_{orig} :
Z'_{syn}  
\end{bmatrix}
\begin{bmatrix}
-\hat{p}_1 \\
1-\hat{p}_2 
\end{bmatrix}
=0
\end{equation}
where $\hat{p_1}$ and $\hat{p_2}$ are vectors of length $n_1$ and $n_2$ of the propensity scores for the original and synthetic data respectively. Thus the first equation corresponding to the intercept gives the following expression for the mean of the propensity score as 
\begin{equation} 
\bar{\hat{p}}= n_2/(n_1+n_2)= n_2/N = c.
\end{equation}

The assumption that $(\hat{p}-c) << (I-c)$, for every row of $Z$ leads to the folllowing results. Since $I$ takes the values 1 and 0 only it follows that all elements of $w$ can be approximated by $c(1-c)$.  This approximation means that we can express the deviation of $logit(\hat{p})$ from its mean as $w^{-1}(\hat{p}-c)$ since the derivative of $logit(\hat{p})$ at its mean is $w^{-1}$. Thus from  equation~(\ref{eq1}) we get
\begin{equation}
Z\hat{\beta}- \bar{Z\hat{\beta}}=Z(Z'Z)^{-1}Z'w^{-1}[(\hat{p}-c)] \\
\end{equation}
which can become
\begin{equation}
Z\hat{\beta}- \bar{Z\hat{\beta}}=Z(Z'Z)^{-1}Z'w^{-1}[(\hat{p}-c) + (I-\hat{p})], \\
\end{equation}
because $W^{-1}(I-\hat{p})$ is zero at convergence as we saw in equation~(\ref{eq1}), and then
\begin{equation}
Z\hat{\beta}- \bar{Z\hat{\beta}}=Z(Z'Z)^{-1}Z'w^{-1}[(I-c)]. \\
\end{equation}
 We can write this in terms of its component matrices as
\begin{equation}
\label{eq10}
Z\hat{\beta}- \bar{Z\hat{\beta}}=
\begin{bmatrix}
Z_{orig} \\
Z_{syn}  
\end{bmatrix}
\begin{bmatrix}
Z'_{orig} Z_{orig}+ Z'_{syn} Z_{syn} 
\end{bmatrix}
^{-1}{}
\begin{bmatrix}
Z_{orig} :
Z_{syn}  
\end{bmatrix}
w^{-1}
\begin{bmatrix}
-n_2/N \\
n_1/N  
\end{bmatrix}
\end{equation} 
where the final column vector consists of a unit vector with $n_1$ entries equal to $-n_2/N$ and $n_2$ entries equal to $n_1/N$. Using the approximation
\begin{equation}
\hat{p} - c = (Z\hat{\beta}- Z\bar{\hat{\beta}}) \frac{d\hat{p}}{d(Z\hat{\beta})}|_{\hat{p}=c}
\end{equation}
we get $\hat{p} - c = (Z\hat{\beta}- Z\bar{\hat{\beta}})w$, since the derivative becomes $c(1-c) = w$. Applying this with equation~(\ref{eq10}) we get
\begin{equation}
\hat{p} - c =
\begin{bmatrix}
Z_{orig} \\
Z_{syn}  
\end{bmatrix}
\begin{bmatrix}
Z'_{orig} Z_{orig}+ Z'_{syn} Z_{syn} 
\end{bmatrix}
^{-1}
\begin{bmatrix}
\bar{Z}_{syn}-\bar{Z}_{orig}
\end{bmatrix}
n_1 n_2/N
\end{equation}
and the mean-squared error from the propensity score becomes 
\begin{equation}
\label{eq20}
pMSE = (\hat{p} - c)' (\hat{p} - c)/N =
\begin{bmatrix}
\bar{Z'}_{syn}-\bar{Z'}_{orig}
\end{bmatrix}
\begin{bmatrix}
Z'_{orig} Z_{orig}+ Z'_{syn} Z_{syn} 
\end{bmatrix}
^{-1}
\begin{bmatrix}
\bar{Z}_{syn}-\bar{Z}_{orig}
\end{bmatrix}
(n_1 n_2/N)^2/N.
\end{equation}
The first element of $[\bar{Z}_{syn}-\bar{Z}_{orig}]$, from the intercept term of the regression, becomes zero, and the expectation of the matrix
\begin{equation}
\begin{bmatrix}
Z'_{orig} Z_{orig}+ Z'_{syn} Z_{syn} 
\end{bmatrix}
^{-1}
\end{equation}
without its first row and column becomes 
\begin{equation}
\begin{bmatrix}
( Z'_{orig}-\bar{Z'}_{orig} )  ( Z_{orig}-\bar{Z}_{orig} )  +
( Z'_{syn}-\bar{Z'}_{syn} )  ( Z_{syn}-\bar{Z}_{syn} )
\end{bmatrix}
^{-1},
\end{equation}
because the independence of $Z_{orig}$ and $Z_{syn}$ ensure that the contribution of cross-product terms to the inverse will be zero. 
When the synthetic data are generated from the distribution that generated $Z_{orig}$ the  expected value of  
$[\bar{Z}_{syn}-\bar{Z}_{orig}] $
will converge to zero for large samples and its variance to $V/n_2$ where $V$ is the variance of $Z_{orig}$.  Also, the expression
\begin{equation}
\label{eq21}
\begin{bmatrix}
( Z'_{orig}-\bar{Z'}_{orig} )  ( Z_{orig}-\bar{Z}_{orig} )  +
( Z'_{syn}-\bar{Z'}_{syn} )  ( Z_{syn}-\bar{Z}_{syn} )
\end{bmatrix}
\end{equation}
will converge to $NV$ for large samples. Thus we can see that equation~(\ref{eq20}) is a multiple of a quadratic form in $\bar{Z}_{orig}-\bar{Z}_{syn}$ of dimension $(k-1)$, so it is distributed as 
\[
\frac{(\frac{n_1}{N})^2(\frac{n_2}{N})}{N}\chi^2_{k-1}.
\]
Thus the expected value and standard deviation of the $pMSE$
\[
E[pMSE]= \frac{(1-c)^2c}{N}(k-1)
\]
\[
 StDev(pMSE)= \frac{(1-c)^2c}{N}\sqrt{(2(k-1)}
\]
becoming $(k-1)/(8N)$ and $\sqrt{2(k-1)}/(8N)$ when $n_1$ and $n_2$ are equal.

The assumption that  $(\hat{p}-c) << (I-c)$ appears to be a rather strong one, but we are not able to derive our results by relaxing it in any way. The assumption is required for the null distribution when we would not expect the regression to provide much discrimination, so, except for small samples,  we would expect all the predicted values to be close to $c$ in that case. The simulation results in Section~\ref{sec:app_sim1} help to confirm that the assumption may be reasonable.

A crucial assumption in this derivation is that the large sample expectation of each column of $Z_{syn}$, under repeated syntheses from the same original data, will be the mean of the corresponding column of  $Z_{orig}$. This follows trivially, and without the asymptotic assumption, for the columns of  $Z_{orig}$ and  $Z_{syn}$ that correspond to $Y_{orig}$. For other columns we note that the expectation of any function of the variables in a distribution can be written as a function of its parameters $\theta$ and that any function of consistent estimators is a consistent estimator of the corresponding function of $\theta$. Thus, for large samples, the means of the columns of  $Z_{orig}$ will be functions of $\hat{\theta}$. Since the columns of  $Z_{syn}$ are combinations of variables generated from $f(y|\hat{\theta})$, their expectation will be given by the same function of $\hat{\theta}$ that defines the mean of the corresponding column of  $Z_{orig}$.

\subsection{Distribution of \textit{\textbf{pMSE}} when some variables are left unchanged}
\label{sec:app_part}
The derivations above require that all elements of each variable are replaced by synthetic values.  When synthesis is incomplete because only some variables are synthesized, while others remain as in the original data, a variant of this result can be used as follows. Some of the predictors $Z$ used in the logistic regression will use only the unsynthesized variables. We can denote this subset by $Z^{fix}$ and the remaining variables by $Z^{*}$ which will be assumed to have $k^{*}$ columns, including the intercept term. The values of  $[\bar{Z}^{fix}_{syn}-\bar{Z}^{fix}_{orig}]$ will all be identically zero because their synthesized values are identical to the original. Thus the $pMSE$ can be written in terms of a quadratic form from $[\bar{Z^*}_{syn}-\bar{Z^*}_{orig}]$.  Thus, by arguments paralleling those above, the $pMSE$ for this type of incompletely synthesized data will have the distribution given above with $k^*$ replacing $k$, where $k^*$ is the dimension of the predictor matrix involving only synthesized variables (including interactions between synthesized and unsynthesized variables). 

\subsection{Distribution of \textit{\textbf{pMSE}} calculated from two synthetic datasets from the same original data}
\label{sec:app_pairs}
When the $pMSE$ is calculated from two synthetic datasets, each synthesisied from the same original data, the expression $[\bar{Z}_{syn}-\bar{Z}_{orig}] $ is replaced by $[\bar{Z}_{syn1}-\bar{Z}_{syn2}] $  which can be written as $[(\bar{Z}_{syn1}-\bar{Z}_{orig}) - (\bar{Z}_{syn1}-\bar{Z}_{orig})] $. Conditional on $Z_{orig}$, these two terms are independent and each has variance covariance matrix $V$. 
Thus the $pMSE$ calculated from two synthesized datasets, each of size $n_2$ will have expected value and standard deviation  $2(k-1)(1-c)^2c/N$, twice that of the null expected $pMSE$ for a sample of size $n_2$. This result was confirmed by simulations from logistic models and applies when the variation between syntheses is due to the differences in the $Z$ matrices between syntheses. However, when CART methods are used we violate some of the theoretical assumptions and thus cannot rely on theoretical expectations. As we will see in the simulations in the next section, the variation between syntheses is only a minor component of the between synthesis variation which is dominated by the differences between the final CART models selected by the algorithms applied to different $Z$ matrices. Such differences are the same between pairs as between the original data and a single synthesis. Thus for CART methods the mean between-pair differences estimate the null $pMSE$.

\section{Appendix II}
\subsection{Simulations of Null \textit{\textbf{pMSE}} with CART Propensity Score Models}
\label{app:resampSim}
Here we show results for simulations using the same setup as described in Section~\ref{sec:app_sim1}, but we switch the logistic propensity score models for non-parametric CART models. In the first case all the data are synthesized, and in the second case only two of the ten variables are synthesized, as before. We estimate the null as described in Section~\ref{sec:resamp} by resampling and comparing pairs
of synthetic datasets which we know were drawn from the same generative model. We carried out corresponding simulations using the permutation method and obtained similar results for complete data like Table \ref{tabA1}. 

\begin{table}[ht]
\caption{\label{tabA1}  Results from 1,000 simulated complete syntheses of multivariate Normal data,  using correct and incorrect models with the $pMSE$ calculated from non-parametric CART models.}
\centering
\begin{tabular}{rrrrrrrr}
  \hline
 Population & \vline & \multicolumn{3}{c}{ Correct Synthesis } & \multicolumn{3}{c}{Incorrect Synthesis } \\ 	
 		Covariance & \vline & \multicolumn{3}{c}{$pMSE$} & \multicolumn{3}{c}{$pMSE$} \\
 		& \vline & Mean & Ratio & Std. & Mean &  Ratio & Std.\\ 
  \hline
  0.0 & \vline & 0.02134 & 0.96584 & -0.20497 & 0.02194 & 0.99360 & -0.04966 \\ 
  0.1 & \vline & 0.02162 & 0.98190 & -0.11288 & 0.02812 & 1.27088 & 1.45820 \\ 
  0.2 & \vline & 0.02134 & 0.97505 & -0.14870 & 0.03800 & 1.71642 & 3.88034 \\ 
  0.3 & \vline & 0.02067 & 0.95014 & -0.27136 & 0.05137 & 2.32482 & 7.11181 \\ 
  0.4 & \vline & 0.02055 & 0.95326 & -0.25137 & 0.06837 & 3.08777 & 11.31911 \\ 
  0.5 & \vline & 0.02051 & 0.95946 & -0.21945 & 0.08866 & 4.01128 & 16.16698 \\ 
  0.6 & \vline & 0.01991 & 0.94439 & -0.28992 & 0.11269 & 5.10356 & 22.04270 \\ 
  0.7 & \vline & 0.01944 & 0.93813 & -0.30129 & 0.14155 & 6.40335 & 29.09392 \\ 
  0.8 & \vline & 0.01893 & 0.93915 & -0.28086 & 0.17431 & 7.89067 & 37.06338 \\ 
  0.9 & \vline & 0.01748 & 0.92756 & -0.26846 & 0.20365 & 9.23024 & 43.92087 \\   
   \hline
\end{tabular}
\begin{tablenotes}
	\footnotesize
\item Ratios and standardized scores calculated from the null distribution estimated from 45 pairs formed from 10 multiple syntheses of each simulated data set.
\end{tablenotes}
\end{table}

Table \ref{tabA2} shows the simulation results for the CART propensity score models. Note that the $pMSE$s calculated for CART models are much larger than those from the logistic models, with the estimated null indicating that differences have been introduced via model-selection resulting in what we might term an over-fitting component of variation. We can see that in this case, as with the simulations using a parametric model, the ratio and standardized $pMSE$ values stay constant for the correct synthesis across different correlations. As can be seen from the first column, the expected value under different correct synthesis changes, since the number of parameters in the propensity score model is no longer fixed. We also see a slight upwards bias in the null, resulting in slight downward bias of the ratio statistics, though it is quite small and should not affect practical use. Another important difference from using the logistic propensity score model is that the ratios for the wrong synthesis increase on a slower scale, which is due to an increase in the $pMSE$ scores both for the observed and the null due to the size of the trees. All of this amounts to a tradeoff in return for the greater flexibility to assess more complex datasets when using CART models.

Relating to what was discussed in Section~\ref{sec:resamp}, it is important to not use overly large trees, since it will make it harder to discern between worse syntheses. As the number of splits (effective parameters) in the tree approaches the number of observations, the expected $pMSE$ ratio will be limited at 2, no matter how bad the synthesis. This is because the maximum value for the $pMSE$ is 0.25, and the maximum expectation under the null is 0.125 (1/8) when the original and synthetic datasets have the same number of rows. When using CART propensity score models, it is important to keep an eye on how many splits there are with respect to the number of observations to get an idea of how the ratio values will scale.

Next, we replicate the incomplete synthesis simulation from Section~\ref{sec:app_sim1} using the CART propensity score model and null approximation. The results shown in Table~\ref{tabA3} are once again what we expect, exhibiting the same patterns as the complete data simulation, and showing that this resampling method can be used to estimate the null $pMSE$ for CART models when not all variables are synthesized.

\begin{table}[ht]
\caption{\label{tabA3} 	 Results from 1,000 simulated syntheses of multivariate Normal data with only two of the 10 columns , synthesized using correct and incorrect models with the $pMSE$ calculated from non-parametric CART models.}
\centering
\begin{tabular}{rrrrrrrr}
  \hline
 Population & \vline & \multicolumn{3}{c}{ Correct Synthesis } & \multicolumn{3}{c}{Incorrect Synthesis } \\ 	
 		Covariance & \vline & \multicolumn{3}{c}{$pMSE$} & \multicolumn{3}{c}{$pMSE$} \\
 		& \vline & Mean & Ratio & Std. & Mean &  Ratio & Std.\\ 
  \hline
  0.0 & \vline & 0.016106 & 0.991741 & -0.028395 & 0.016259 & 0.987100 & 0.004145 \\ 
  0.1 & \vline & 0.015956 & 0.996262 & -0.012367 & 0.019528 & 1.207032 & 0.738865 \\ 
  0.2 & \vline & 0.015331 & 0.993628 & -0.019922 & 0.024894 & 1.534692 & 1.912159 \\ 
  0.3 & \vline & 0.015085 & 1.012724 & 0.038179 & 0.031738 & 1.967164 & 3.618090 \\ 
  0.4 & \vline & 0.013968 & 0.998023 & -0.005400 & 0.040502 & 2.559000 & 5.430248 \\ 
  0.5 & \vline & 0.012879 & 0.989164 & -0.028306 & 0.051799 & 3.290095 & 7.915648 \\ 
  0.6 & \vline & 0.012119 & 1.004021 & 0.009555 & 0.066470 & 4.217474 & 11.584360 \\ 
  0.7 & \vline & 0.010523 & 1.002788 & 0.005776 & 0.085163 & 5.488950 & 16.148351 \\ 
  0.8 & \vline & 0.009099 & 0.989744 & -0.019050 & 0.110642 & 7.366080 & 20.769889 \\ 
  0.9 & \vline & 0.007091 & 1.054471 & 0.087836 & 0.148356 & 10.417990 & 29.718231 \\ 
   \hline
\end{tabular}
  	\begin{tablenotes}
	\footnotesize
	\item Ratios and standardized scores calculated from the null distribution estimated from 45 pairs formed from 10 multiple syntheses of each simulated data set.
\end{tablenotes}
\end{table}

\clearpage
\bibliographystyle{chicago}
\bibliography{synthCombined}

\end{document}